\documentstyle[preprint,aps]{revtex}
\begin{document}
\draft
\preprint{}
\title{Exact Schwarzschild-Like Solution for SU(N) Gauge Theory}
\author{D. Singleton}
\address{Department of Physics, University of Virginia,
Charlottesville, VA 22901}
\date{\today}
\maketitle
\begin{abstract}
In this paper we extend our previously discovered exact solution
for an SU(2) gauge theory coupled to a massless, non-interacting
scalar field, to the general group SU(N+1).
Using the first-order formalism of
Bogomolny, an exact, spherically symmetric solution for the
gauge and scalar fields is found. This solution is similiar
to the Schwarzschild solution of general relativity, in that
the gauge and scalar fields become infinite at a radius,
$r_0 = K$, from the origin. It is speculated that this may be the
confinement mechanism that has long been sought for in non-Abelian
gauge theories, since any particle which carries the SU(N+1) charge
would become permanently trapped once it entered the region $r <
r_0$. The energy of the field configuration of this solution is
calculated.
\end{abstract}
\newpage
\narrowtext
\section{The SU(N+1) Schwarzschild-like Solution}

In a previous paper \cite{sing} we exploited the connection between
general relativity and Yang-Mills theory to find an exact
Schwarzschild-like solution for an SU(2) gauge theory coupled to
a massless scalar field. In the present paper we wish to show that
a similiar solution can be found for the general group SU(N+1).
Instead of using the Euler-Lagrange formalism which leads to coupled
second-order, nonlinear equations, we will use the Bogomolny
approach \cite{bogo} to derive our field equations.
Bogomolny  obtained his first-order version of the Yang-Mills
field equations by requiring that the gauge and scalar fields
produce an extremum of the canonical Hamiltonian. The field equations
obtained in this way are first-order, but their solutions are
also solutions to the second-order Euler-Lagrange equations.

The model which we consider here is an SU(N+1) gauge field coupled
to a massless scalar field in the adjoint representation. The
Lagrangian for this theory is
\begin{equation}
\label{lagran}
{\cal L} = -{1 \over 4} F^{\mu \nu a} F_{\mu \nu} ^a + {1 \over 2}
D^{\mu} ( \Phi ^a ) D_{\mu} ( \Phi ^a)
\end{equation}
where
\begin{equation}
F_{\mu \nu} ^a = \partial _{\mu} W_{\nu} ^a - \partial _{\nu}
W_{\mu} ^a + g f ^{abc} W_{\mu} ^b W_{\nu} ^c
\end{equation}
and
\begin{equation}
D_{\mu} \Phi ^a = \partial _{\mu} \Phi ^a + g f ^{abc}
W_{\mu} ^b \Phi ^c
\end{equation}
where $f^{abc}$ are the structure constants of the gauge group
and $a, b, c = 1, 2, \dots , N, N+1$.
The canonical Hamiltonian obtained from Eq. (\ref{lagran}) is
\begin{equation}
\label{hamo}
{\cal H} = \int d^3 x \left[{1 \over 4} F_{ij} ^a F^{aij} - {1 \over 2}
F_{0i} ^a F^{a0i} + {1 \over 2} D_i \Phi ^a D^i \Phi ^a
- {1 \over 2} D_0 \Phi ^a D^0 \Phi ^a \right]
\end{equation}
We wish to find gauge and scalar fields which produce
an extremum of ${\cal H}$.
First we rescale the scalar field ({\it i.e.} $\Phi ^a \rightarrow
A \Phi ^a$). This is done so that later on it will be simple to examine
the pure gauge case by setting $A=0$.
Next we specify that all the fields are time independent, and that
the time component of the gauge fields are proportional to the
scalar fields ({\it i.e} $W_0 ^a = C \Phi ^a$, where $\Phi ^a$ is
the rescaled scalar field). The time component of the gauge fields
act like an additional Higgs field except its kinetic term appears
with the opposite sign in the Lagrangian \cite{zee}. Using these two
requirements and the antisymmetry of $f^{abc}$ we find that
$D_0 \Phi ^a = 0$ and $F_{0i} ^a = C (D_i \Phi ^a)$, so that the
Hamiltonian becomes
\begin{eqnarray}
\label{ham}
{\cal H} = \int d^3 x \Big[&&{1 \over 4} \left(F_{ij} ^a - \epsilon_{ijk}
\sqrt{A^2 - C^2} D^k \Phi ^a \right) \left(F^{aij} - \epsilon_{ijl}
\sqrt{A^2 - C^2} D^l \Phi ^a \right) \nonumber \\
+ && {1 \over 2} \epsilon_{ijk} \sqrt{A^2 - C^2}
F^{aij} D^k \Phi ^a \Big]
\end{eqnarray}
Using the fact that
\begin{equation}
{1 \over 2} \epsilon_{ijk} F^{aij} D^k \Phi ^a = \partial ^i
\left({1 \over 2} \epsilon_{ijk} F^{ajk} \Phi ^a \right)
\end{equation}
and the requirement that the solutions we are looking for are
only functions of $r$ we find
\begin{eqnarray}
\label{ham1}
{\cal H} &=& \sqrt{A^2 - C^2} \int _S (\Phi^a B^a _i)
dS ^i \nonumber \\
&+& \int d^3 x \left[{1 \over 4} \left(F_{ij} ^a -
\epsilon_{ijk} \sqrt{A^2 - C^2} D^k \Phi ^a \right)
\left(F^{aij} - \epsilon_{ijl} \sqrt{A^2 - C^2}
D^l \Phi ^a \right) \right]
\end{eqnarray}
For the total divergence term we
have used the definition of the non-Abelian
magnetic field in terms of the field strength tensor ({\it i.e}
$B^a _i = {1 \over 2} \epsilon_{ijk} F^{ajk}$), and used
Gauss's Law to turn the volume integral into a surface
integral. The lower limit of this Hamiltonian can be found by
requiring
\begin{eqnarray}
\label{fbogo}
F_{ij} ^a &=& \epsilon _{ijk} \sqrt{ A^2 - C^2}
D^k \Phi ^a \nonumber \\
or \nonumber \\
B_i ^a &=& \sqrt{A^2 - C^2} D_i \Phi ^a
\end{eqnarray}
To get the second expression we have again used the definition of
the non-Abelian magnetic field. These are the Bogomolny equations
\cite{bogo}. Wilkinson and Goldhaber \cite{gold} have given a
generalized ansatz for the gauge and scalar fields
\begin{eqnarray}
\label{ansatz}
{\bf W}_i &=& {\epsilon_{ijb} r^j ({\bf T}^b - {\bf M}^b (r)) \over g r^2}
\nonumber \\
\Phi ^a &=& {{\bf \Phi} (r) \over g}
\end{eqnarray}
${\bf W} _i$ are three $(N+1) \times (N+1)$ matrices of the gauge
fields. ${\bf M}_b (r)$ and ${\bf \Phi} (r)$ are four $(N+1) \times
(N+1)$ matrices whose elements are functions of $r$, and in terms of
which the Bogomolny equations will be written. ${\bf T}_b$ are three
$(N+1) \times (N+1)$ matrices which generate the maximal embedding
of SU(2) in SU(N+1). Because of the spherical symmetry requirement
one can look at Eq. (\ref{fbogo}) along any axis \cite{bais}
\cite{wilk}. Taking the positive $\hat {z}$ axis the Bogomolny
equations become $\sqrt{ A^2 - C^2} (D_3 \Phi ^a) = B _3 ^a$ and
$\sqrt{A^2 - C^2} (D_{\pm} \Phi ^a) = B_{\pm} ^a$, or in terms of the
ansatz of Eq. (\ref{ansatz})
\begin{eqnarray}
\label{fbogo1}
r^2 \sqrt{A^2 - C^2} {d {\bf \Phi} \over dr} &=&
[{\bf M}_+  , {\bf M}_- ] - {\bf T}_3  \nonumber \\
{d {\bf M}_{\pm} \over dr} &=& \mp \sqrt{A^2 - C^2}
[{\bf M}_{\pm} , {\bf \Phi} ]
\end{eqnarray}
Taking the third ``component'' of the maximal SU(2) embbedding into
SU(N+1) as
\begin{equation}
{\bf T}_3 = diag \left[{1 \over 2}N, {1\over 2}N -1, \dots ,
-{1 \over 2} N +1, -{1 \over 2} N \right]
\end{equation}
it has been shown \cite{gold} that the matrix functions, ${\bf M}_+ (r)$
and ${\bf \Phi} (r)$, can be taken as
\begin{eqnarray}
\label{mat1}
{\bf \Phi} = {1 \over 2}
\left(
\begin{array}{ccccc}
\phi_1 &\space &\space &\space &\space \\
\space &\phi_2 - \phi_1 &\space &\space &\space \\
\space &\space &\ddots &\space &\space \\
\space &\space &\space &\phi_N -\phi_{N-1} &\space \\
\space &\space &\space &\space &- \phi _N \\
\end{array}
\right)
\end{eqnarray}
\begin{eqnarray}
\label{mat2}
{\bf M}_+ = {1 \over \sqrt{ 2}}
\left(
\begin{array}{ccccc}
0 &a_1 &\space &\space &\space \\
\space &0 &a_2 &\space &\space \\
\space &\space &\ddots &\space &\space \\
\space &\space &\space &0 &a_N \\
\space &\space &\space &\space &0 \\
\end{array}
\right)
\end{eqnarray}
where $\phi _m$ and $a_m$ are real functions of $r$ and ${\bf M}_- =
({\bf M} _+) ^T$. Substituting Eqs. (\ref{mat1}), (\ref{mat2}) into
the first order field equations of Eq. (\ref{fbogo1}) the field
equations become \cite{bais} \cite{wilk}
\begin{eqnarray}
\label{deqn}
r^2  {d \phi _m \over dr} &=& {1 \over \sqrt{A^2 - C^2}}
\left[ (a_m)^2 - m{\bar m} \right] \nonumber \\
{d a_m \over dr} &=& \sqrt{A^2 - C^2} \left(-{1 \over 2} \phi_{m-1}
+\phi_m - {1 \over 2} \phi _{m+1} \right) a_m
\end{eqnarray}
where $1 \le m \le N$, ${\bar m} = N +1 - m$ and $\phi_0 = \phi_{N+1} =0$.
Exact solutions have been found to Eq. (\ref{deqn}) \cite{bais}
\cite{wilk} which are generalizations of the well known
Prasad-Sommerfield solution \cite{prasad} for SU(2). The
Prasad-Sommerfield solution and their generalizations satisfy
the boundary condition that the gauge and scalar fields are
finite at the origin. If one does not require that the fields
be finite at the origin then the coupled equations for $\phi _m (r)$
and $a_m (r)$ can be solved by
\begin{eqnarray}
\label{soln}
\phi _m (r) &=& {1 \over \sqrt{A^2 - C^2}}
{K m {\bar m} \over r (K - r)} \nonumber \\
a_m (r) &=& {r \sqrt{m {\bar m}} \over K - r}
\end{eqnarray}
$K$ is an arbitrary constant with the dimensions
of distance. This is the generalization of a similiar
solution which we found for SU(2) using the second-order
Euler-Lagrange formalism \cite{sing}. The reason for wanting to
generalize our solution to SU(N+1) is to give a possible explanation for
the confinement mechanism in QCD whose gauge group is SU(3). Inserting
the functions $\phi _m (r)$ and $a_m (r)$ of Eq. (\ref{soln}) into
the fields of Eq. (\ref{ansatz}) we find that these fields become
infinite at a finite radius of
\begin{equation}
r_0 = K
\end{equation}
The non-Abelian ``electric'' ($E_i ^a = F_{0i} ^a$) and ``magnetic''
fields ($B_i ^a = {1 \over 2} \epsilon_{ijk} F^{jka}$) calculated for
this solution also become infinite at $r_0$. Thus any particle which
carries an SU(N+1) charge would either never be able to penetrate beyond
$r_0$ (when the SU(N+1) charges are replusive) or once it passed into
the region $r < r_0$ it would  never be able to escape  back to the
region $r > r_0$. One has in a sense a color charge black hole.

Both the present solution and our SU(2) solution were found by
using the connection between Yang-Mills theory and general relativity
\cite{utiyama}, and trying to find the Yang-Mills equivalent of the
Schwarzschild solution. The objects in general relativity which correspond
to the gauge fields are the Christoffel coefficients, $\Gamma^{\alpha}
_{\beta \gamma}$. Examining a few of the Christoffel symbols of the
Schwarzschild solution we find
\begin{eqnarray}
\Gamma ^t _{r t} &=& {2GM  \over 2r ( r - 2GM )} \nonumber \\
\Gamma ^r _{r r} &=& -{2GM  \over 2r ( r - 2GM )}
\end{eqnarray}
where $2GM$ the equivalent of the constant $K$ from the Yang-Mills
solution. The similarity between these Christoffel coefficients and the
gauge and scalar fields that result from the solutions, $\phi _m (r)$
and $a_m (r)$ of Eq. (\ref{soln}), is striking. The most important
similarity from the point of explaining confinement is the existence
in both solutions of an event horizon, from which particles which carry
the appropriate charge can not escape once they pass into the region
$r < r_0$. For general relativity the appropriate ``charge'' is
mass-energy so that nothing can climb back out of the Schwarzschild
horizon, while in the Yang-Mills case only particles carrying an SU(N+1)
charge will become confined.

One slightly disturbing feature of these Schwarzschild-like SU(N+1)
solutions is that they have an infinite energy due to the
singularity at $r = 0$.
When quantities such as the energy of this field configuration
are calculated the integral must be cutoff at some
arbitrary radius, $r_c$. The singularity at
$r_0$ does not give an infinite energy unless one sets
$r_c = K$. This singular behaviour at the origin is shared
by several other classical field theory solutions. Both the
Schwarzschild solution of general relativity, and the Coulomb solution
in electromagnetism have similiar singularities.  The Wu-Yang
solution \cite{wu} for static SU(2) gauge fields with no time component
also blows up at the origin, leading to an infinite energy if one
integrates the energy density down to $r=0$. Just as these classical
solutions are not expected to hold down to $r=0$ so the present solution
will certainly be modified by quantum corrections as $r$ approaches
zero. Phenomenologically we know that the present solutions can not
be correct for very small $r$, since they do not exhibit the asymptotic
freedom behaviour that is a desirable consequences of the quantum
corrections to QCD. It would be interesting to see if the behaviour of
the fields at the origin could be modified by introducing a mass
term (${m^2 \over 2} \Phi ^a \Phi ^a$) and a self interaction term
(${\lambda \over 4} (\Phi ^a \Phi ^a) ^2$) to the scalar field part of
the Lagrangian, while still retaining the color event horizon feature of
the present solution. This smoothing of the fields at the origin does
happen when one compares the Prasad-Sommerfield exact solution (where
there are no mass or self interaction terms for the scalar field) with
the the numerical results of 't Hooft \cite{thooft} or Julia and Zee
\cite{zee} (where mass and self interaction terms are included).
The numerical results lead to monopoles and dyons with a
finite core, while the exact Prasad-Sommerfield solution leads to a
point monopole (despite this their exact solution still has finite
energy when the energy density is integrated down to zero, unlike our
present solution). As in the case of the Prasad-Sommerfield solution,
introducing mass and self interaction terms for the scalar fields
would require solving the equations numerically, since we have
not been able to find an analytical solution under these conditions.

To calculate the energy of the field configuration of our solution it is
necessary to integrate the $T_{00}$ component of the energy-momentum
tensor  over all space, excluding the origin. The  $T_{00}$
component of the energy-momentum tensor
is similiar to the Hamiltonian density of Eq. (\ref{ham}) except that
all the terms have positive signs
\begin{equation}
T_{00} = {1 \over 4} F_{ij} ^a F^{aij} + {1 \over 2} F_{0i} ^a
F^{a0i} + {A^2 \over 2} D_i \Phi ^a D^i \Phi ^a  + {A^2 \over 2}
D_0 \Phi ^a D^0 \Phi ^a
\end{equation}
The energy in the fields of our solution is the integral over all
space of $T_{00}$. Using the field equations of Eq. (\ref{fbogo}),
and the fact that $F_{0i} = C(D_i \Phi ^a)$ the energy of the fields
is
\begin{eqnarray}
E &=& \int T_{00} d^3 x \nonumber \\
&=& A^2 \int d^3 x D_i \Phi ^a D^i \Phi ^a
\end{eqnarray}
Since the fields only have a radial dependence the angular part
of the integration can be easily done. Futher using the radial
symmetry to evaluate the integrand along the positive $\hat{ z}$
axis, and using the matrix expression for $\Phi ^a$ as well as the
field equations, Eq. (\ref{fbogo1}), we find that the energy
becomes \cite{gold}
\begin{equation}
\label{energy1}
E = {8 \pi A^2 \over g^2} \int _{r_c} ^{\infty} r^2 dr \left(
Tr \left( \left[{d {\bf \Phi } \over dr} \right] ^2 \right)
+ {2 \over \ A^2 - C^2} Tr \left( r^{-2} {d {\bf M_+} \over dr}
{d {\bf M_-} \over dr} \right) \right)
\end{equation}
Using the solutions for the elements of the matrices,
$\bf{ \Phi}$ and $\bf{ M}_{\pm}$ of Eq. (\ref{soln}) we find
\begin{eqnarray}
\label{phim}
Tr \left( \left[ {d {\bf \Phi } \over dr} \right]^2 \right)
&=& {K^2 (2r - K)^2 \over 4 r^4 (A^2 - C^2)
(K - r)^4} \sum _{n=0} ^N (N - 2n)^2 \nonumber \\
Tr \left( r^{-2} {d {\bf M_+} \over dr} {d {\bf M_-} \over dr} \right)
&=& {K^2 \over r^2 (K-r)^4} \sum _{n=0} ^N n(N+1-n)
\end{eqnarray}
Using this in Eq. (\ref{energy1}) and carrying through the integration
the energy in the field configuration of this Schwarzschild-like
solution is
\begin{equation}
\label{energy2}
E = {2 \pi A^2 K^2 N(N+1)(N+2) \over 3 g^2 (A^2 - C^2)}
\left[ {K  - 2 r_c \over r_c (K -r_c)^3} \right]
\end{equation}
where the sums from Eq. (\ref{phim}) have been done explicitly.
This result can be checked against the SU(2) result \cite{sing}
by taking  $N=1$ in Eq. (\ref{energy2}), and the expressions for
the energy do indeed agree (In Ref. \cite{sing} we required that
$A^2 - C^2 =1$ whereas here this factor is divided out). As it
stands there are two arbitrary constants that enter the solution
({\it i.e.} $K$ and $r_c$) which would have to be specified
before any connection between this Schwarzschild-like solution
and the real world could be carried out. As has already been
mentioned $K$ is the Yang-Mills equivalent of $2GM$ in general
relativity. Thus it could be conjectured that $K$ is related
to the strength of the gauge interaction
($G$ in general relativity), and the
magnitude of the central charge which produces the gauge field
configuration ($M$ in general relativity).

One interesting feature which this general SU(N+1) Schwarzschild-like
solution shares with our previous SU(2) solution is that scalar
fields are apparently required in order to get a physically
non-trivial solution. If there where no scalar fields in the original
Lagrangian ({\it i.e.} $A = 0$), then the field energy of Eq.
(\ref{energy2}) would be zero, and the $W_0 ^a$ component of the
gauge fields would be pure imaginary. Although the pure gauge
case with no scalar fields is a solution mathematically, its
physical significance is dubious. Requiring that the
solutions are pure real, or that the energy in the fields
be non-zero would exclude the pure gauge case solution. Under either
of these requirements on the solution, it can be seen that scalar
fields must be present.

\section{Conclusions}

In this paper we have generalized our previous exact,
Schwarzshild-like solution for SU(2) Yang-Mills theory
to SU(N+1) by using an embedding of SU(2) into SU(N+1) \cite{gold}.
This exact SU(N+1) solution was found by using the connection
between general relativity and Yang-Mills theories. It was found that
the Schwarzschild solution of general relativity carries over
with only a little modification into an equivalent solution for an
SU(N+1) gauge theory coupled to massless scalar fields. Just as the
Schwarzschild solution possesses an event horizon which permanently
confines any particle which carries the ``charge'' of the  gravitational
interaction ({\it i.e} mass-energy), so the present solution also
has a ``color'' event horizon which permanently confines any particle
which carries the SU(N+1) gauge charge. This may be the confinement
mechanism which has long been sought for the SU(3) gauge theory of
the strong interaction, QCD. Before this claim can be made there
are several important questions which must be resolved. First, under
some reasonable physical assumptions about the nature of our solution
it is found that scalar fields are required for a solution to
exist. Normally scalar fields are not thought to play a significant role
in confinement, so the physical importance of these scalar fields
would need to be addressed. Second, there are several arbitrary constants
which crop up in the solution ($K$ and $r_c$). In order to make a
connection with the real world these constants would have to be given.
Theoretically $K$ should be related to the strength of the gauge
interaction as well as the magnitude of the gauge charge which produces
the Schwarzschild-like gauge fields. Experimentally $K$ should be
related to the radius of the various QCD bound states ({\it e.g.}
protons, pions, etc.). The other constant, $r_c$, was
introduced chiefly to avoid the
singularity at $r=0$, but also because our solution does not
possess the property of asymptotic freedom as $r \rightarrow 0$.
This should not be too surprising since our solution is
for classical Yang-Mills fields, but as $r \rightarrow 0$ quantum
effects should become increasingly important.
Thus $r_c$ can be thought of roughly, as marking the boundary between
the classical, confining solution of this paper, and the quantum
dominated asymptotic freedom regime. All this strongly suggests
a bag-like structure for QCD bound states : As a particle approaches
$r = K$ from $r < K$, it feels a progressively  stronger color force
which confines it to remain inside the bound state. As the particle
approaches $r \rightarrow 0$ it enters the asymptotic freedom
regime, where it moves as if it were free.

An interesting extension of this work would be to see if other
exact solutions from general relativity have Yang-Mills counterparts.
In particular if a Yang-Mills equivalent of the Kerr solution
could be found it might give some insight into the nature of
the spin of fermions.

\section{Acknowledgements}  I would like to acknowledge the help
and suggestions of David Singleton and Hannelore Roscher.

\end{document}